\documentclass[aps, 12pt, tightenlines]{revtex4-2}
\usepackage{graphicx}
\usepackage{amsmath}
\usepackage{amsfonts}

\bibstyle{apsrev4-2}

\usepackage[colorlinks=true, allcolors=blue]{hyperref}
\usepackage{url}

\begin{document}
	
\title{Monolithic bowtie cavity traps for ultra-cold gases.}

\author{Yanping Cai}
\author{Daniel G. Allman}
\author{Jesse Evans}
\author{Parth Sabharwal}
\author{Kevin C. Wright}

\affiliation{Dartmouth Department of Physics and Astronomy, Hanover, NH, 03755, USA}
\email[Correspondence may be directed to: ]{ kevin.wright@dartmouth.edu}

\begin{abstract}
We report on trapping and cooling $^6$Li atoms in a monolithic ring bowtie cavity. To make the cavity insensitive to magnetic fields used to tune atomic interactions, we constructed it entirely from fused silica and Zerodur. The components were assembled using hydroxide bonding, which we show can be compatible with ultra-high vacuum. Backscattering in high-finesse ring cavities readily causes trap intensity fluctuations and heating, but with phase-controlled bi-directional pumping the trap lifetime can be made long enough for quantum gas experiments in both the crossed-beam trap (unidirectional pump) and 2D lattice trap (bidirectional pump) configurations.
\end{abstract}

\maketitle

\section{Introduction}

Optical dipole traps are an essential tool for experiments with ultracold atoms and molecules. Efficient capture is more difficult for atoms with unresolved excited state hyperfine structure (Li, K), buffer-cooled gases, and Stark-decelerated molecules which are at relatively higher temperatures. Trapping atoms in the field of a resonantly pumped optical cavity circumvents some of the problems associated with control of high-power optical traps, allowing the creation of large and deep traps using less pump laser power~\cite{Mosk2001}. There is an extensive history of trapping and cooling atoms in optical cavities~\cite{Ritsch2013} which we will not review in depth. Most experiments have used linear (two-mirror) cavities to create a standing-wave dipole trap (1D optical lattice) where the atoms are usually be transferred to a single site or secondary trap for efficient evaporative cooling~\cite{Mosk2001, Weimer2014}.

Few ultra-cold atom experiments have used cavities with three or more mirrors~\cite{KruseColdPRA03, Bernon2011, Weimer2014, Culver2016, Megyeri2018, Naik2018, Wen2020}. They are more complex, but the ability to operate the cavity in standing-wave or traveling-wave mode~\cite{Nagorny2003, Weimer2014} provides greater flexibility for separately optimizing capture, cooling, and other experimental objectives. One important consideration for evaporative cooling in cavity-based traps is that evaporation in a crossed-beam trap is more efficient than in an elongated single-beam trap~\cite{Olson2013}. Four or more mirrors are required to form a self-intersecting cavity mode, and bosonic atoms have been trapped~\cite{Bernon2011} and cooled to degeneracy~\cite{Naik2018} in this configuration. Ultra-cold fermionic atoms ($^6$Li) have previously been trapped in a linear cavity~\cite{Mosk2001}, trapped in one arm of a bidirectionally pumped ring bowtie cavity~\cite{Weimer2014}, and cooled to degeneracy in a linear cavity~\cite{Roux2020}, but this work is the first time fermionic atoms have been trapped at the cavity self-intersection point.

\begin{figure}[ht]
	\centering
	\includegraphics[width=5.5in]{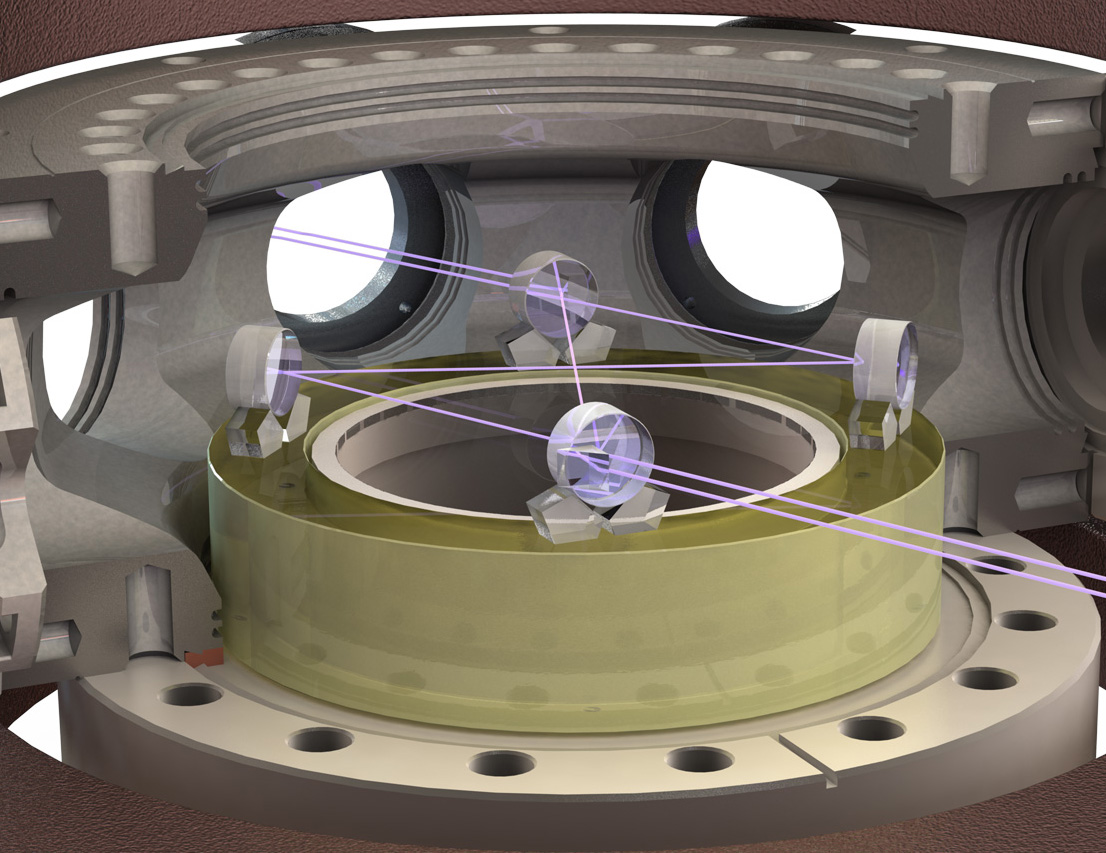}
	\caption{Cut-away rendering of the monolithic ring bowtie cavity inside our lithium quantum gas apparatus. Four 12.7 mm diameter cavity mirrors are supported by two fused silica prisms each, and by a ring-shaped Zerodur spacer~\cite{bondoptics}. Two of the input/output beam paths are redirected through the vacuum viewports by right-angle prism mirrors bonded to the back of the cavity coupling mirrors.}
	\label{fig:cav_rend}
\end{figure}

We ultimately constructed a monolithic cavity from fused silica and Zerodur in a symmetric ring-bowtie configuration (See Fig.~\ref{fig:cav_rend}). Because the cavity contains no conductive materials it is unaffected by rapid changes in the magnetic fields used for the magneto-optical trap and for controlling interactions between the lithium atoms. The current cavity is far-detuned and singly-resonant, but we expect to use what we have learned to design a doubly-resonant tunable monolithic cavity with a sub-recoil ($<$60 kHz) linewidth for light near resonance with the lithium D lines. We note that the strong collective coupling limit was recently achieved with a unitary gas of $^6$Li atoms in a linear cavity~\cite{Roux2020}, and fermions with tunable interactions in a ring cavity could provide new opportunities for studying interesting aspects of collective light-matter interactions~\cite{SteinkeRolePRA11} including collective self-organization, superradiant emission and collective atomic recoil lasing~\cite{KruseColdPRA03, Robb2004, Bonifacio2005, Fallani2005}.  Cavity-cooling techniques are of great fundamental and practical importance~\cite{Aspelmeyer2014}, and there are numerous untested predictions about cooling mechanisms for (esp. fermionic) atoms in a ring cavity~\cite{Elsasser2003, Sandner2011, Wolke2012, Sandner2013, Colella2019, Gietka2019}. Atom-ring-cavity interactions can also generate dynamical optical potentials~\cite{Ritsch2013,KruseColdPRA03}, artificial gauge fields, and spin-orbit coupling ~\cite{Mivehvar2015}. We expect there may also be interesting applications for such a cavity in developing atom-optical gyroscopes, accelerometers, or gradiometers.

\section{Mechanical Design}

The experimental demands of a quantum gas experiment impose many constraints on the cavity design. The first is that the cavity mirrors and supports must allow optical access for laser cooling, imaging, etc. Typical beams for a magneto-optical trap are 15-25 mm in diameter, and high numerical aperture imaging requires the largest possible unobstructed view along at least one axis. We fit the cavity within our vacuum chamber by mounting the 12.7 mm diameter mirrors on a ring-shaped piece of class 2 Zerodur glass ceramic that occupies an otherwise unused space in the lower part of the vacuum chamber (see Fig.~\ref{fig:cav_rend}). The inside (outside) diameter of the spacer is 78 mm (106 mm)~\cite{bondoptics}. The height of the spacer is 25 mm, and its top surface is at the same height as the upper surface of the recessed viewport in the bottom of the chamber. 

Many quantum gas experiments require large, rapid changes to magnetic fields, and a high-finesse optical cavity can easily be disturbed by these changing fields. Standard vacuum-compatible optical mounts are usually made of austenitic stainless steel, which is weakly magnetic. Aluminum and titanium are non-magnetic, but forces from eddy currents can still disturb a cavity with sub-MHz linewidth. The best way to eliminate magnetic forces on the cavity is to use completely non-conductive materials. A monolithic cavity is also much more rigid and stable than a comparable assembly using metal kinematic mounts; the thermal stability of a bonded fused silica or glass ceramic assembly will be at least an order of magnitude better than any metallic structure of similar dimensions. The trade-off is that the cavity alignment is fixed, and the cavity length is only weakly adjustable via temperature tuning. Extremely high precision is required in the alignment and bonding process to ensure that the cavity beams focus and intersect accurately at the center of symmetry, since any error is permanent.

It is also essential for the cavity to be vacuum compatible, and be able to survive the bake-out conditions required to achieve UHV/XHV. The materials the cavity is composed of (fused silica and Zerodur) already meet this requirement. The bonds between components must also not be a source of outgassing, or have virtual leaks from trapped contaminants. We placed our cavity on four small fluoroelastomer (Viton) o-rings with a square profile (ID: 0.75mm, OD 2.75 mm, thickness 1 mm), after removing a 30 degree sector from the ring to avoid creating a virtual leak. Pre-baking fluoroelastomer can reduce the outgassing rate to below 1$\times10^{-10}$ Torr$\,\cdot\,$L$\,\cdot\,$s$^{-1}\,\cdot\,$cm$^{-1}$\cite{Peacock1980}. For the total exposed surface area of 20 mm$^2$ we expected that the gas load from these small pieces of elastomer would not significantly affect the vacuum. With the cavity installed on these supports our vacuum-limited lifetime is around 30 seconds.

The small fluoroelastomer supports provide primary isolation and damping for the cavity, and the chamber is mounted on a large vibration-isolated optical table. The water-cooled electromagnets surrounding the chamber are mounted on a highly damped composite support structure that has no connection with the vacuum chamber, except through the optical table. Light tapping on the vacuum chamber produces a measurable disturbance of the cavity lock, but tapping on the coil mounting assembly does not, and there is no observable change in the cavity noise spectrum when the cooling water flow is turned on.

The lowest-frequency mechanical modes of the cavity are torsional and translational center-of-mass modes with frequencies around 90 Hz. The next group of mechanical modes are center-of-mass tipping and bouncing modes with frequencies around 500 Hz. The first vibrational mode affecting the relative positions and angles of the mirrors is a radial quadrupole mode at 5 kHz. This and all similar vibrational modes are in the multi-kHz frequency range due to the stiffness of the solid Zerodur ring spacer. This large frequency gap between center of mass modes and vibrational modes helps reduce mode coupling and decreases the sensitivity of the cavity resonance frequency to external mechanical disturbances.

\begin{figure}[ht]
	\centering
	\includegraphics[width=5in]{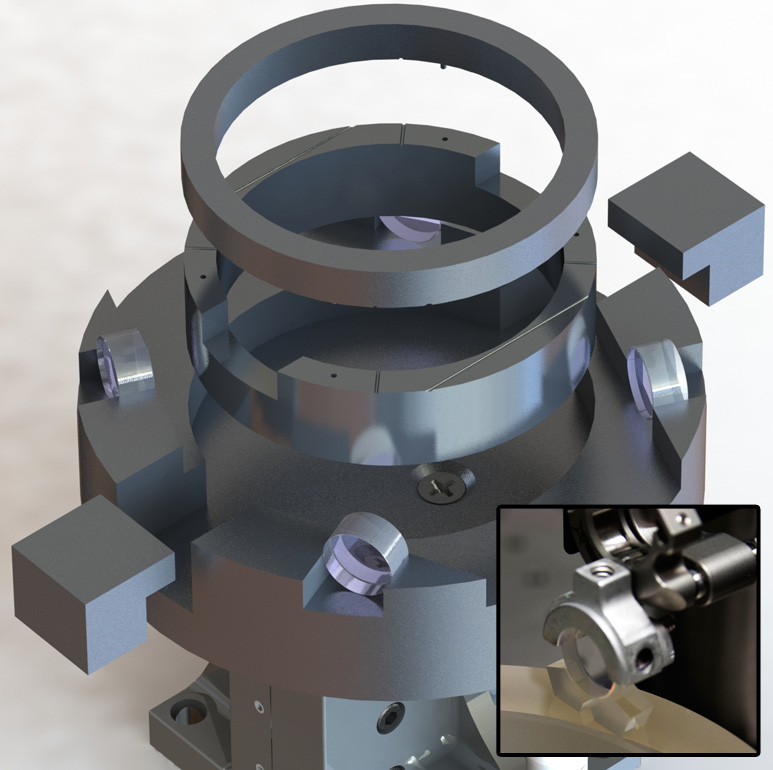}
	\caption{Illustration of mechanical elements used in coarse and fine alignment. The largest element in the exploded view of the jig is the primary mechanical reference, which is lowered on a translation stage to allow fine alignment once the mirrors are held by a kinematic assembly, shown in the inset. The grooved cylindrical inserts ensure that the cavity remains square and flat during fine alignment. The inset photograph shows one of the cavity mirrors in its fine-alignment mount just after bonding of the support prisms. }
	\label{bjwi}
\end{figure}
 
\section{Assembly Methods}

To assemble the cavity we used hydroxide bonding, which was previously used to create monolithic optical systems for LIGO and the LISA Pathfinder missions~\cite{Elliffe2005}. The bond curing rate is slow enough to allow fine alignment, and the fully cured bonds are stronger and more reliable than optical contacting. We found no published information on the suitability of hydroxide bonding for use in UHV environments, so we tested numerous cavity prototypes with different mirrors, substrates, and bonding solutions. We found that using 1-2 $\mu$L of a 3:1 dilution of sodium silicate solution resulted in dense UHV-compatible bonds that survived bake-out to at least 150 C and any typical (non-impact) stresses. Bond strength is affected by the geometry, roughness, and wetting properties of the surfaces, but with sodium silicate solution the process is robust enough to bond fine-ground glass surfaces with different surface figures.

The tight positioning tolerances for the cavity required a fairly complex alignment procedure. We used a mechanical reference to assist with the alignment process (See Fig.~\ref{bjwi}). One critical part of that reference was a removable two-part insert with guide grooves of 140 $\mu$m radius machined along the intended path of light in the cavity. With the top of the guide removed, the mode-matched pump beam was aligned to the guide grooves along one side of the cavity then. We then offset the pump beam horizontally by 1.7 mm to account for walk-off of the beam when passing through the input coupling mirror (6.35 mm thick fused silica at an angle of 22.5 degrees). We mounted the cavity mirrors into D-shaped holders and placed the mirrors in slots in the jig (12.7 mm wide, 6.35 mm deep) which constrained their positions with an estimated accuracy of around 0.1 mm. The axial position of each mirror was set by sliding it forward until its edge came into contact with a high-precision gauge block temporarily placed against a reference face machined into the jig.

After coarse mechanical positioning we were already able to see the cavity transmission signal on a photodetector and identify the TEM$_{00}$ mode. For fine alignment, we bonded a compact five-axis kinematic assembly to the back of each mirror holder with a small drop of epoxy. We then lowered the mechanical reference by 2 mm to allow sufficient clearance for fine alignment. The vertical position and gap between the guide grooves were adjusted with metal shims, and we sequentially adjusted mirrors $B$, $C$, $D$, and $A$, in that order (See Fig.~\ref{fig:cavity_schematic}). We reduced the vertical gap between the grooves slightly after each iteration until it was fully closed, and then we again maximized the transmission of the TEM$_{00}$ cavity mode.

At the end of the fine alignment procedure we verified that the beams intersected at the cavity center by moving a sharp tungsten (STM) tip around that point with a three-axis translation stage and measuring the decrease in cavity transmission. We found that the beams were coplanar over an 8 mm square area to within the 2 $\mu$m resolution of the stage, well within our target tolerance of 10 $\mu$m. We then removed the mechanical reference and placed the Zerodur ring under the mirrors with its top 10.8 mm below the plane of the cavity, so the cavity would be in the vertical mid-plane of the experimental chamber. We positioned two fused silica penta prisms (5 mm) under the sides of each mirror, bonded them to the Zerodur, and then bonded the mirrors to the prisms. (See inset in Fig.~\ref{bjwi}). Finally, we bonded two small right-angle prism mirrors to the back of the cavity coupling mirrors so that light reflected off the cavity would be redirected back out through the chamber viewports instead of striking the chamber wall (See Fig.~\ref{fig:cav_rend}).
	
Hydroxide bonds take several days to reach full strength, but with our test cavities the bonds sometimes failed if the mirrors were left in the fine-positioning assembly for that long. Assembly was performed in a laminar flow hood on a vibration-isolated optical table in a space temperature controlled to 0.1$^\circ$C, but we suspect that differential thermal expansion or spurious mechanical shock caused those bond failures. We found we could prevent bond failure by removing the kinematic assembly as soon as the bonds were strong enough to withstand the forces exerted during the removal process (typically 2 hours). We monitored the cavity transmission signal and saw no indication that the mirrors shifted during removal of the fine positioning assembly. Additional measurements with the STM tip confirmed that the cavity alignment was unchanged. We allowed 2 days for the bonds to reach full strength and performed a final UV/ozone cleaning \cite{HansenUltravioletOzoneApplOpt93} before installing the cavity in the vacuum chamber. 

\section{Optical Design}
\label{sec:opticaldesign}

The most common laser choice for optical dipole traps is a Yb-doped fiber laser operating around 1064 nm.  The Stark shift of the $2S\rightarrow2P$ (671 nm)  transitions in lithium at this wavelength is $6\times10^{-4}$ Hz/(W/m$^2$)\cite{Burchianti2014}. For a trap depth of 1 mK the shift is around 9 MHz (1.5 $\Gamma$), which is small enough that it does not significantly affect laser cooling~\cite{Grier2013}. It is also possible to use the $2S\rightarrow3P$ (323 nm) transitions for narrow-line laser cooling~\cite{DuarteAllPRA11}, and the ``magic'' trap wavelength where the Stark shift vanishes for this transition happens to be 1068.8 nm. The marginal cost of a laser operating at that wavelength rather than 1064 nm was negligible, and so we chose to operate our cavity at 1068.8 nm to leave open the option of future experiments utilizing the narrow-line cooling transitions.

The optical design of the cavity is significantly constrained by the physical layout of the vacuum chamber and the need to provide access for the MOT beams, the cold atomic beam from the 2DMOT, transfer to the science cell, magnetic coils, and imaging systems. The cavity mirrors are 12.7 mm in diameter, and a side length $s = 59$ mm allows adequate clearance for other beams to pass (see Fig.~\ref{fig:cav_rend}). The cavity round-trip distance is $d = 290$ mm, resulting in a cavity free spectral range (FSR) of 1.03 GHz. 

\begin{figure}[ht]
	\centering
	\includegraphics[width=4.5in]{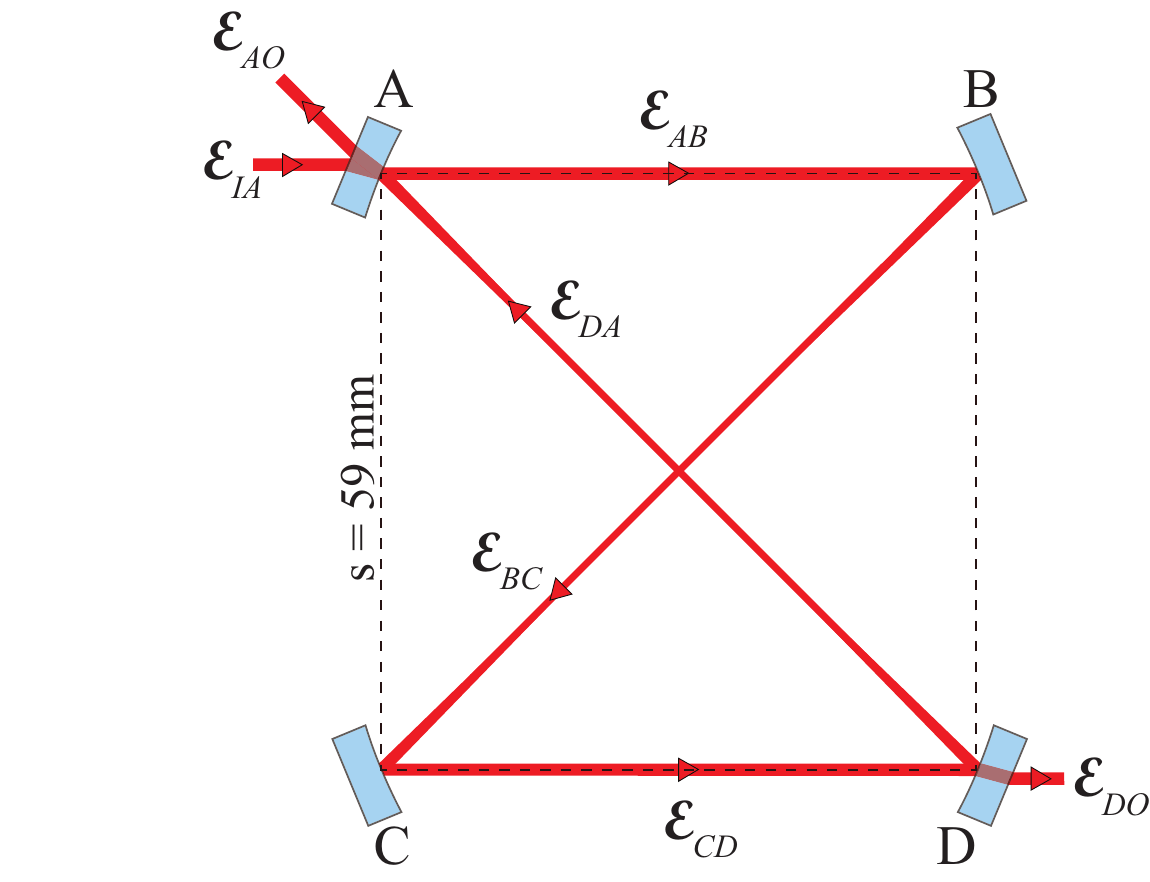}
	\caption{Schematic depicting our four-mirror symmetric bowtie cavity. The mirrors labeled A and D are the input/output couplers, and mirrors B and C are high-reflectors. }
	\label{fig:cavity_schematic}
\end{figure}

To derive expressions for important properties of the cavity we assign labels $A$,$B$,$C$,$D$ to the cavity mirrors, and use the symbols $\mathcal{E}_{xx'}$ to distinguish the different input, output and intracavity fields as shown in Fig.~\ref{fig:cavity_schematic}. The amplitude reflection and transmission coefficients of mirror $x$ are $r_x$ and $t_x$ respectively. We can then use the relations between these quantities to derive general expressions for the cavity amplitude reflection and transmission coefficients $\mathbf{r}_x$ and $\mathbf{t}_x$ for a single pump field entering mirror $A$, in terms of the mirror coefficients
\begin{align}
\mathbf{t}_B &= \frac{\mathcal{E}_{BO}}{\mathcal{E}_{IA}} = \frac{t_A t_B \exp{(ikd_{AB}})}{1 - \tilde{r}\exp{(ikd)}}\\
\mathbf{t}_C &= \frac{\mathcal{E}_{CO}}{\mathcal{E}_{IA}} = \frac{ t_A r_B t_C \exp{(ikd_{ABC})}}{1 - \tilde{r}\exp{(ikd)}}\\
\mathbf{t}_D &= \frac{\mathcal{E}_{DO}}{\mathcal{E}_{IA}} = \frac{ t_A  r_B r_C t_D \exp{(ikd_{ABCD})}}{1 - \tilde{r}\exp{(ikd)}}\\
\mathbf{r}_A &= \frac{\mathcal{E}_{AO}}{\mathcal{E}_{IA}} = \frac{t_A^2 r_B r_C r_D \exp{(ikd)}}{1 - \tilde{r} \exp{(ikd)}}  - r_A
\end{align}
Here the cavity round-trip return amplitude is $\tilde{r} = r_A r_B r_C r_D$ and $d_{AB...}$ is the part of the cavity distance between mirrors $A$, $B$, etc. The cavity reflectance and transmittance are

\begin{align}
\mathbf{T}_B &= \frac{t_A^2 t_B^2 / (1-\tilde{r})^2 }{1 + F(\tilde{r})\sin^2(kd/2)}\\
\mathbf{T}_C &= \frac{t_A^2  r_B^2 t_C^2/(1-\tilde{r})^2}{1 + F(\tilde{r})\sin^2(kd/2)}\\
\mathbf{T}_D &= \frac{ t_A^2 r_B^2 r_C^2 t_D^2/(1-\tilde{r})^2}{1 + F(\tilde{r})\sin^2(kd/2)}\\
\mathbf{R}_A &= r_A^2\frac{(1-\tilde{r}\epsilon_A)^2 \left(1 + F(\tilde{r}\epsilon_A)\sin^2(kd/2)\right)}{(1-\tilde{r})^2 \left(1 + F(\tilde{r})\sin^2(kd/2)\right)}\label{eq:RA}
\end{align} 
where $\epsilon_A=(r_A^2+t_A^2)/r_A^2=(1-L_A)/r_A^2$ is a parameter expressing the effect of having a power loss $L_A$ on the input coupling mirror $A$, and we have generalized the usual definition of the coefficient of finesse to be a function:
\begin{equation}
F(r)=\frac{4 r}{(1-r)^2}
\end{equation}
The finesse of the cavity is related to the mirror reflectivities through the coefficient of Finesse in the usual way:
\begin{equation}
\mathcal{F}(\tilde{r})=\frac{\pi \sqrt{F(\tilde{r})}}{2}=\frac{\pi \sqrt{\tilde{r}}}{1-\tilde{r}}
\end{equation}

The choice of mirror reflectivities also affects the pump beam input coupling efficiency $\mathbf{C}= 1 - \mathbf{R}$. Perfect input coupling efficiency is only possible when the cavity reflectance vanishes. From Eq.\eqref{eq:RA} we can see that this will only occur on resonance, and only when the mirror reflectivities satisfy the constraint $1-\tilde{r}\epsilon_A = 0$. This can be written more explicitly in terms of the properties of the individual mirrors:
\begin{equation}
r_A=r_B r_C r_D (1-L_A)\label{eq:impedancematch}
\end{equation}
Perfect input coupling is not possible when all four mirrors have identical reflectivities, even when there is no loss. In that case, even on resonance 25\% of the incident power is reflected~\cite{Bernon2011}. One way to satisfy Eq.~\eqref{eq:impedancematch} is to use one mirror with lower reflectivity ($r_L$) as an input coupler, and choose a higher reflectivity ($r_H$) for the other three mirrors, where $r_L=r_H^3(1-L_A)$. It is also possible to use two lower-reflectivity mirrors as coupling ports, with the remaining two mirrors having a high reflectivity $r_H\approx1$. If $r_H$ is high enough, the cavity impedance is essentially the same as a cavity with two identical mirrors. The cavity we built uses two input coupler mirrors and two high-reflector mirrors with the specifications given in Table~\ref{table:mirrorspec}. It should have a finesse above $2\times10^4$, a linewidth less than 50 kHz, and a coupling efficiency greater than 90\%.

\setlength{\tabcolsep}{0.5em} 
\begin{table}[ht]
	\caption{Manufacturer's specifications for reflectance ($R$), transmittance ($T$), and loss ($\kappa$) of our cavity mirrors for $\lambda=1070$ nm, at a 22.5$^\circ$ angle of incidence.  Values listed are in parts per million. Subscripts $p$ and $s$ indicate polarization parallel and perpendicular to the plane of the cavity.}
\centering
	\begin{tabular}{ | c || p{0.9cm} | p{0.6cm} || p{0.9cm} | p{0.6cm} || p{0.6cm} | }
		\hline
		& $1-R_p$  & $T_p$ & $1-R_s$ &  $T_s$ & $\kappa$\\
		\hline\hline
		Coupler Mirrors &  100 & 70 & 40 & 10 & 30\\
		\hline
		High Reflectors &  40 & 25 & 20 & 5 & 15\\
		\hline
	\end{tabular}
	\label{table:mirrorspec}
\end{table}

If the cavity mirrors are spaced as shown in Fig.~\ref{fig:cavity_schematic} and they all have the same radius of curvature, $R_\mathrm{m}$, there are two different ranges of $R_\mathrm{m}$ for which the cavity is stable~\cite{Abitan2005}. For 40 mm $< R_\mathrm{m} <$ 60 mm, the smallest beam waists occur in the short sides of the cavity, which is not optimal for trapping at the center. For $R_\mathrm{m}>100$ mm, the minimum spot size occurs at the cavity center-of-symmetry, so this is the only range we will consider. Within this range, the cavity could be described as quasi-concentric when $R_\mathrm{m} \approx 100$ mm, and quasi-confocal for $R_\mathrm{m} \approx 250$ mm. A concentric cavity has the smallest possible beam waist at the center and the greatest trap depth for fixed optical power, but it also has the smallest capture volume. Increasing the mirror radius closer to the confocal range has several effects. First, increasing $R_m$ reduces the astigmatism of the cavity, making it easier to match the pump beam to the cavity spatial mode. The capture volume of the trap increases, which is advantageous as long as the trap depth can be maintained by increasing the intracavity power. 

Other experiments where atoms have been trapped at the self-intersection point of a ring bowtie cavity~\cite{Bernon2011, Naik2018} have operated close to the quasi-concentric (tight-focus) limit, and the overlap of the beams at the center could be adjusted. We were very concerned about ensuring good beam overlap at the center of the cavity during the process of aligning and bonding our cavity, since any error would be permanent. We ultimately selected mirrors with $R_m = 200$ mm, making the cavity closer to confocal than concentric, and relaxing the tolerances on beam overlap. At our operating wavelength of 1068.8 nm, the $1/e^2$ radius of the cavity mode at the focus is $w_\parallel = 150$ $\mu$m in the plane of the cavity, and $w_\perp = 159$ $\mu$m perpendicular to the cavity. We chose to pump the cavity by injecting into the side rather than the diagonal because the cavity mode is closer to being symmetric and collimated in that segment of the cavity, which makes mode matching less difficult. Given this beam geometry, the capture volume of the central intersection region is around 10 times larger than most crossed beam traps (assuming a typical beam radius of $50$ $\mu$m). For an intracavity power $P$ (in Watts), the trap depth for lithium atoms is 3.2 $\mu K \times P$; 300 Watts is required to create a 1 mK deep trap. The trap frequencies for lithium scale with power as follows: $\nu_\parallel = \sqrt{P}$ 450 Hz and $\nu_\perp = \sqrt{P}$ 600 Hz (with $P$ in Watts).

At $R_m = 200$ mm the higher order mode spacings are slightly less than half the FSR, so the cavity is close to confocal. The cavity is slightly astigmatic, with the horizontal transverse mode spacing (440 MHz) is slightly less than the vertical mode spacing (490 MHz). The difference in phase delay upon reflection breaks the degeneracy of modes polarized parallel and perpendicular to the plane of the cavity by 28 MHz. This is smaller than the transverse mode frequency shifts, but much larger than the cavity linewidth. We were able to trap atoms pumping the cavity with either polarization, but all information about trapping presented below is for pumping with the polarization in the plane of the cavity.

\setlength{\tabcolsep}{0.5em} 
\begin{table}
\centering
	\caption{Measured cavity linewidth and finesse for both polarizations compared against calculations for a best-case total cavity loss of $\kappa_c=50$ ppm, and a more realistic cavity loss of 550 ppm. Frequencies are in kHz.}
	\begin{tabular}{ | c || c | c || c | }
		\hline
		& $\kappa_c$=50 ppm & $\kappa_c$=550 ppm & Measured \\
		\hline\hline
		$\Delta\nu_\mathrm{FWHM}^{(p)}$ &  40 & 134 & 130(10)\\
		\hline
		$\mathcal{F}^{(p)}$&  $2.5\times10^4$ & $8\times10^3$ & $8\times10^3$ \\
		\hline
		\hline
		$\Delta\nu_\mathrm{FWHM}^{(s)}$&  15 & 90 &  100(10)\\
		\hline
		$\mathcal{F}^{(s)}$ &  $7\times10^4$ & $1.1\times10^4$ & $1\times10^4$ \\
		\hline
	\end{tabular}\label{table:cavityspec}
\end{table}

The mirrors we used to construct the cavity~\cite{layertec} had magnetron sputtered coatings, which can have losses as low as 10 ppm.  State-of-the-art ion-beam sputtered coatings can have losses of $<1$ ppm, leaving considerable room for improved performance in possible future experiments. The manufacturer's loss specifications for our mirrors (see Table~\ref{table:mirrorspec} ) were somewhat larger than the state-of-the-art for magnetron sputtered coatings, and we expected to observe a total cavity loss of at least 50 ppm due to mirror imperfections.

 We took precautions to avoid mirror contamination during assembly and installation, but the measured performance of the cavity after bake-out indicated total losses of about 550 ppm (see Table~\ref{table:cavityspec}). The maximum input coupling efficiency was reduced to about 25\%, and heating of the mirrors made it more difficult to operate the cavity at the power level required to capture lithium from a MOT. An intracavity power of 300 Watts corresponds to a 600 kW/cm$^2$ peak intensity on the mirrors, and and the power lost or absorbed at any given mirror could easily exceed $100$ mW.  Previous studies of thermal effects on low-loss mirrors reported that for low losses the radius of curvature changed by 50-100 $\mu$m/(MW/cm$^2$), becoming flatter at higher intensity~\cite{Uehara1995, Meng2005}. Another study with higher absorption losses (1000 ppm) measured much larger changes of 100 mm/(MW/cm$^2$)~\cite{Edmunds2013}. Our cavity losses are around 120 ppm per mirror, so we expect the curvature change falls somewhere in between these two extremes. We observe that above 50 Watts the intracavity power no longer changes linearly with the pump power. Above 100 Watts, the intracavity power becomes hysteretic, and feedback is required to smoothly ramp the power up and down. At 300 Watts and above the intracavity power occasionally jumps spontaneously. In spite of these complications, we do not lose the frequency lock when these jumps occur, and we have operated the cavity at power levels sufficient to capture atoms directly from a laser-cooled cloud of $^6$Li (see Fig.~\ref{fig:atomsincavity}).

\section{Locking and Stability} \label{sec:locking}

 \begin{figure}[ht]
	\centering
	\includegraphics[width=5.5in]{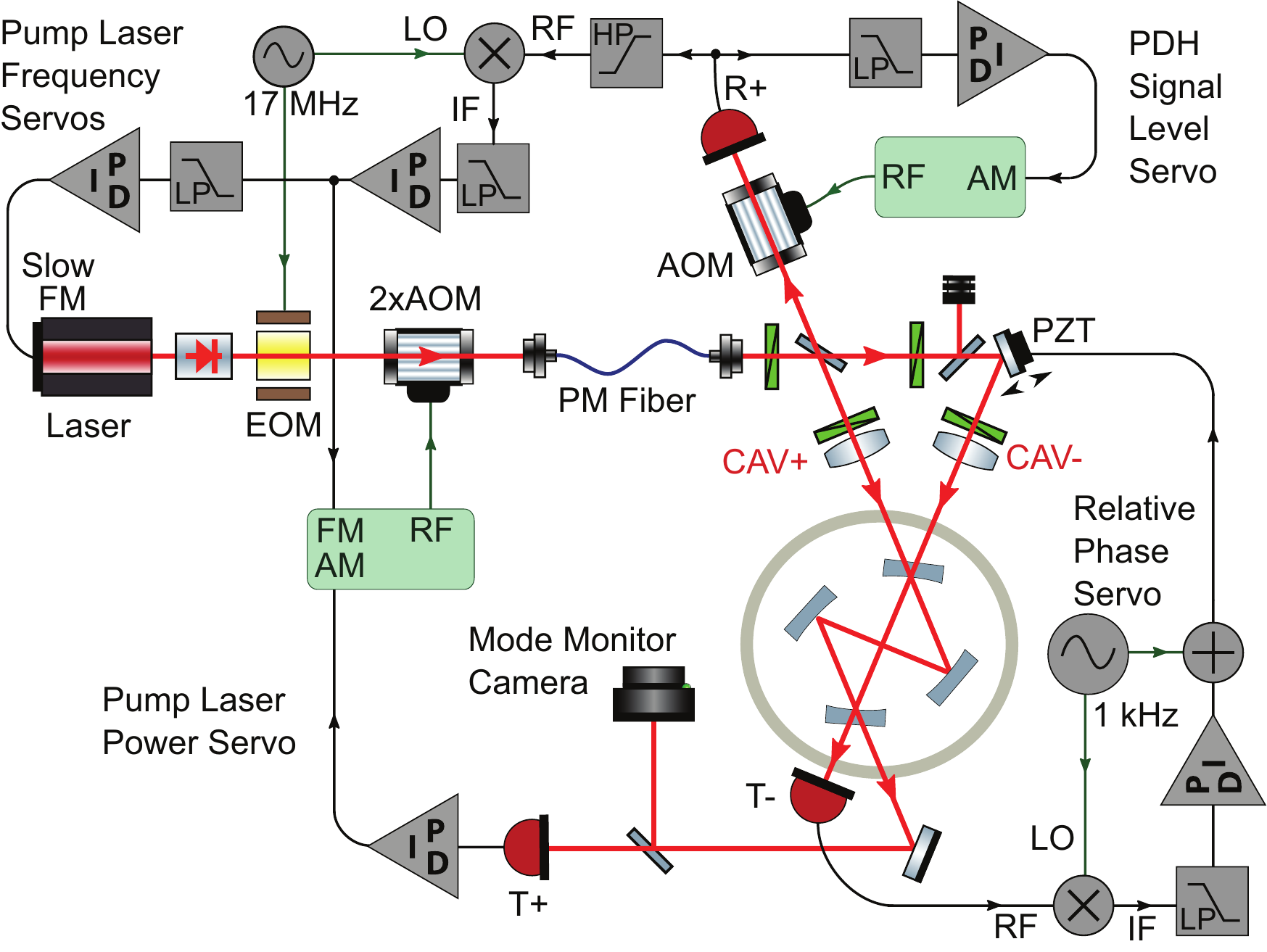}
	\caption{Overview of the locking scheme. The pump laser passes through an EOM which generates 17 MHz sidebands for frequency locking. It then passes through a double-passed AOM (2xAOM) used for overall power and fast frequency control. The common pump is split into into forward (CAV+) and reverse (CAV-) pump beams using polarizers and manually controlled waveplates. CAV- is reflected off a piezo-actuated mirror that imparts a controllable phase shift relative to CAV+. Both beams are mode matched to the cavity and injected on the back of the same coupling mirror in opposing directions. The frequency locking signal is derived from the reflection of CAV+ incident on photodetector R+, which is maintained at a constant signal level by a servo-controlled AOM. The Pound-Drever-Hall signal is then sent to a pair of cascaded servos: a fast servo controlling the frequency shift from the 2xAOM, and a slow drift-correction servo controlling the frequency tuning piezo in the pump laser. Transmitted light from the forward and reverse cavity modes strikes photodiodes T+  and T- respectively. The signal from T+ is used to stabilize the forward mode power via servo control of the (common) pump power using the 2xAOM. T- is used to stabilize the reverse mode power by via servo control of the relative phase of CAV+ and CAV-. }
	\label{fig:revpumpschematic}	
\end{figure}

\begin{figure}[ht]
	\centering
	\includegraphics[width=3in]{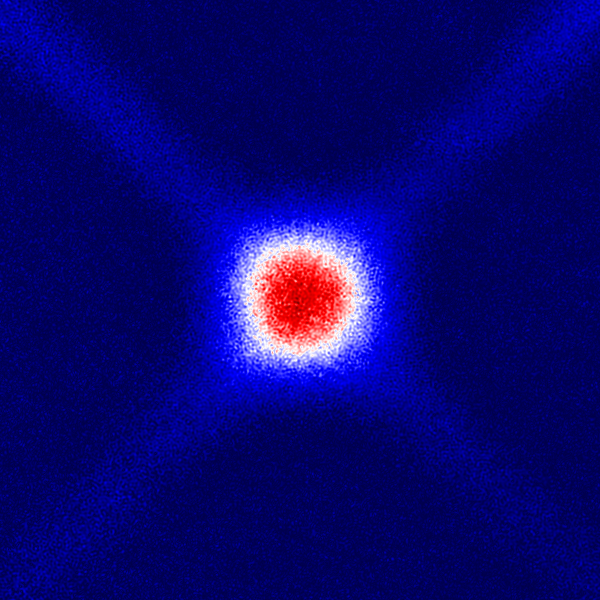}
	\caption{Absorption image of 10$^7$ $^6$Li atoms captured by the cavity trap directly from a gray molasses. The field of view of the image is 1 mm.}
	\label{fig:atomsincavity}
\end{figure}

 The pump laser used in the experiment was a tunable 15 W single frequency MOPA fiber laser system (Koheras Boostik) operating near 1068.8 nm. Coarse wavelength adjustment is possible using temperature control, but was not necessary for these experiments. The signal for locking the frequency of the pump laser to the cavity resonance was obtained using the Pound-Drever-Hall (PDH) technique, with an EOM generating 17 MHz sidebands on the pump at a modulation depth of $<2\%$. The frequency locking signal was obtained from part of the forward pump beam reflected off the cavity injection mirror, as shown in Fig.~\ref{fig:cavity_schematic}. The optical power incident on the detector was stabilized to maintain optimal signal-to-noise as the pump power was changed. The raw PDH signal was demodulated and sent to two cascaded servo loops: a fast loop (200 kHz bandwidth) controlling the frequency shift given to the pump beam by a double-passed 80 MHz AOM, and a slower loop (10 Hz bandwidth) correcting frequency drifts by adjusting the voltage on the tuning piezo of the pump laser. We stabilized the forward cavity mode power by feeding the transmission signal from the other cavity coupling mirror to a servo controlling the power of the same double passed 80 MHz AOM used for fast frequency tuning. With these feedback controls we were able to maintain cavity lock over a wide dynamic range with intensity fluctuations of 0.1\% rms. The manufacturer's specifications for short-term linewidth of the pump laser is less than 10 kHz, but millivolt-level noise on the frequency tuning piezo was enough to cause an RMS frequency deviation on the order of the 100 kHz cavity linewidth, over a time scale of 100 ms. Reductions in piezo noise and/or higher bandwidth in the servo controlling the double-passed AOM should allow further reductions in intensity noise. Even if some of the frequency jitter is due to mechanical excitations instead of pump laser fluctuations, we note that the corresponding RMS path length deviation would be at most 50 pm, given the 1 GHz FSR of the cavity. When loading the trap with $^6$Li atoms captured in a MOT and cooled to 40 $\mu$K by grey-molasses cooling, we typically capture around $10^7$ atoms (See Fig.~\ref{fig:atomsincavity}). About half of these atoms are lost in the first 100 ms after turning off the cooling beams. In our earliest capture attempts we observed a trap lifetime of less than one second, after that rapid initial loss. Servo loop optimization increased the trap lifetime to several seconds, but this was still a factor of 10 below the vacuum-limited lifetime.

We determined that the main remaining cause of the heating of the atoms was fluctuations from backscattered light propagating in the counter-rotating cavity mode. The fractional power varied from 1-10\% of that in the forward-propagating mode, and the amplitude of the field appeared to be more strongly affected by vibrations than the power in the forward mode. Backscatter in a ring cavity is unavoidable due to imperfections in the cavity mirrors; depending on the ratio of loss to coherent backscatter it can even result in coupling and mode splitting of degenerate counter-propagating cavity modes. Low-finesse cavities are more insensitive to backscatter, but in higher-finesse ring cavities minute amounts of backscatter can result in undesirable effects such as mode-locking. Experiments with atoms trapped in high-finesse ring cavities have previously observed backscatter pumping of the counter-propagating mode~\cite{Krenz2007, Bernon2011} with heating and loss caused by fluctuations in the the resulting weak optical lattice. However, more recent experiments in an intersecting ring bowtie have been able to address this heating problem effectively enough to achieve quantum degeneracy with $^{87}$Rb~\cite{Naik2018}.

Backscattering can sometimes be reduced by choosing a longitudinal mode or adjusting the mirror positions so that the phase of the backscattered fields from each mirror cancel~\cite{Slama2007}, but this is not a viable solution for a monolithic cavity. A more effective general solution is to actively control the reverse mode by pumping it with a field of independently controllable amplitude and phase.~\cite{Krenz2007}. Bidirectional pumping capability is desirable in any case, and so we chose to split off part of the power in the forward pump beam to inject into the reverse mode. We reflected this beam off a piezo-mounted mirror at a small angle, allowing high-speed control of the the relative phase of the pump beams (See Fig.~\ref{fig:revpumpschematic}). We found there was little drift in the power required for backscatter cancellation on a one-hour timescale, and manual control of the relative power of the beams using half wave plates and polarizing beamsplitters was sufficient. The phase required for cancellation varied significantly over a timescale of less than a minute, so we modulated the mirror position at a frequency of 1 kHz and used lock-in detection on the reverse mode transmission signal to generate an error signal sent to the servo controlling the mirror position.

With the forward-propagating mode stabilized at 100 W and the counter-injected beam power and phase optimized to fully cancel the backscattered field, the trap lifetime increased to more than 10 seconds. Holding the power in the forward mode stable, we increased the counter-injection beam power and changed the phase lock to stabilize the reverse mode at 50 W. In this configuration the cavity mode forms a 2D optical lattice with trap depth of around 1 mK and transverse trap frequencies in excess of 1 MHz. We observed an increased trap lifetime in this configuration, comparable to the vacuum-limited lifetime measured in a separate crossed-beam dipole trap. We were initially surprised by this result, but suspect that it is a consequence of both the increased trap depth and strong filtering of pump noise at frequencies much higher than the bandwidth of the cavity. 

\section{Conclusion}

We have shown that it is possible to create a bonded all-glass ring cavity that is compatible with the requirements of quantum gas experiments, and trapped ultracold atoms at the self-intersection point of a monolithic symmetric ring bowtie cavity built using these techniques. With servo control of both the forward and reverse pump beams we have achieved trapping times of more than 10 seconds. Adjustment of the pump configuration allows us to load atoms into a crossed-beam trap or 2D optical lattice, and smoothly ramp between those configurations. This is the first realization of a cavity-based 2D optical lattice, and it appears possible to achieve transverse trap frequencies high enough for Raman sideband cooling of lithium in the lattice~\cite{Hamann1998, Parsons2015}. With some further improvements to the servos and locking system it should be possible to cool atoms to degeneracy and conduct quantum gas experiments on atoms confined in the cavity. We are now developing a design for a new doubly resonant cavity (671 nm, 1068 nm) with lower losses. This should allow us to test novel cavity cooling schemes, perform quantum-non-demolition measurements of the collective spin of atoms trapped in the cavity, and study other unusual configurations involving collective coupling between the atoms and the cavity modes. 

\section{Acknowledgments}
We thank Sarah Khatry for work on early cavity prototypes, Dwayne Adams for fabricating the cavity alignment jig, and Roberto Onofrio for valuable discussions and a careful reading of the manuscript.

\bibliography{cavity}

\begin{thebibliography}{40}
\bibitem [{bon()}]{bondoptics}%
  \BibitemOpen
  \href@noop {} {}\bibinfo {note} {The ring spacer was fabricated by Bond
  Optics (Lebanon, NH) from Grade 2 Zerodur. The top surface had a flatness
  specification of $<\lambda/4$, the remaining surfaces had a standard
  ground-glass finish.}\BibitemShut {Stop}%
\bibitem [{lay()}]{layertec}%
  \BibitemOpen
  \href@noop {} {}\bibinfo {note} {The mirrors were fabricated by Layertec GMBH
  (P/N 100337 and 109556)}\BibitemShut {NoStop}%
\end{thebibliography}

\end{document}